\documentclass[twocolumn,showpacs,preprintnumbers,amsmath,amssymb]{revtex4}
\usepackage{graphicx}
\usepackage{dcolumn}
\usepackage{bm}

\begin{document}

\preprint{physics/0605206}

\title{Alpha-muon sticking and chaos in muon-catalysed "in flight" $d$-$t$ fusion} 
\author{Sachie Kimura}
\author{Aldo Bonasera}
\altaffiliation[Also at ]{Libera Universit\`{a} Kore, Enna, Italy.}
\affiliation{Laboratorio Nazionale del Sud, INFN,
via Santa Sofia, 62, 95123 Catania, Italy}
\date{\today}

\begin{abstract}
We discuss the alpha-muon sticking coefficient in the muon-catalysed ``in flight" d-t fusion 
in the framework 
of the Constrained Molecular Dynamics model. Especially the influence of muonic chaotic dynamics 
on the sticking coefficient is brought into focus. 
The chaotic motion of the muon affects not only the fusion cross section but also 
the $\mu-\alpha$ sticking coefficient. Chaotic systems lead to larger enhancements with respect to 
regular systems because of the reduction of the tunneling region. Moreover they give smaller sticking 
probabilities than those of regular events. 
By utilizing a characteristic of the chaotic dynamics one can avoid losing the muon in 
the $\mu$CF cycle.
We propose the application of the so-called ``microwave ionization of a Rydberg atom" to the present case 
which could lead to the enhancement of the reactivation process by using X-rays.   
\end{abstract}

\maketitle

\section{introduction}
The muon catalyzed fusion ($\mu$CF) of hydrogen isotopes, especially 
$d$-$t$ fusion, has been studied as a realizable candidate of an energy source 
at thermal energies.
In the liquid D$_2$ and T$_2$ mixture, the muon assists the fusion through the formation 
of a muonic molecule, since the size of the muonic molecule is much smaller than that of 
the ordinary molecules and the fusing nuclei tend to stay closer.  
After the fusion process the muon is released normally and again it is utilized for 
another fusion.   
The efficiency of the $\mu$CF is governed by the muon-sticking on 
the $\alpha$ particle which is produced in the fusion~\cite{ca,davies}.
The muon is lost from the $\mu$CF cycle due to sticking($\omega_0$), unless it is not released 
successively through the interaction with the medium.    
The rate of the stripping of the stuck muon from the $\alpha$ particle is known as the 
reactivation coefficient $R$ and 
thus the effective sticking probability($\omega_s^{eff}$) is determined by
\begin{equation}
  \label{eq:omeg}
  \omega_s^{eff}=\omega_0(1-R).
\end{equation}
The determination of the value of $R$ is discussed in the reference~\cite{smsw}.
In this paper we do not take into account the medium effects which are supposed to be important 
to determine the precise value of $R$ in the actual experimental setup. We rather aim to propose 
a method in order to enhance the reactivation process, by making use of the stochastic instability
of the stuck muon in an oscillating field. For this purpose we are mainly interested in 
investigating the impact of the regular and chaotic dynamics~\cite{ kb-cdf}. 
The experimental value of the initial sticking $\omega_0$ which is determined assuming 
the theoretical $R$ in~\cite{smsw} is tabulated in the reference~\cite{PhysRevA.49.4481}. 
In the table the values of $\omega_0$ from 7 separate measurements are smaller than theoretical 
estimate for the most part.   
The direct measurement of $\omega_0$ has been conducted as well and gave the 
$\omega_0=0.69\pm0.40\pm0.14$ \%~\cite{davies}.    
The temperature dependent phenomena in the muon cycling rate and 
in the muon loss probability, which is a function of $\omega_s^{eff}$, have been reported  
on the $dd \mu$ at the temperature from 85K to 790K~\cite{bom} 
and on the $dt \mu$ from 5K to 16K~\cite{kawamura1,kawamura2} 
by measuring the fusion neutron yield and the K$_{\alpha}$ X-ray yield lately. They have observed that 
the muon loss probability increases and the muon cycling rate decreases as the temperature decreases.   
In the latter case $\omega_s^{eff}$ varies from 0.64\% to 0.86$\pm$0.01\% as the temperature varies 
from 16K to 5K. 
The temperature dependence in the muon loss probability is thought to be caused by a change of 
the reactivation coefficient $R$ in Eq.~(\ref{eq:omeg})~\cite{kawamura1,kawamura2}. 

At thermal energies, where the $\mu$CF takes place, fluctuations might play a role. 
We investigate the influence of the fluctuations by using 
a semi-classical method, the constrained molecular dynamics (CoMD) approach~\cite{pmb}.
As it is well known the molecular dynamics
contains all possible correlations and fluctuations due to the initial conditions(events). 
In the CoMD, the constraints restrict the phase space configuration of the 
muon to fulfill the Heisenberg uncertainty principle. 
The results are given as an average and a variance over ensembles of 
the quantity of interest which is determined in the simulation. 
Especially we determine the enhancement factor of the reaction cross section by the muon 
as a function of the incident energy in the ``in flight" fusion~\cite{melezhik}: 
\begin{equation}
  t\mu + d \rightarrow (td\mu)^+ \rightarrow {\bf ^4He}^{++} + \mu + n + Q,  \label{eq:dtmm0} \\
\end{equation}
where $Q=$17.59 MeV is the decay Q-value of this reaction.
The enhancement factor of each event indicates the regularity of the system.    
The reaction (\ref{eq:dtmm0}) is known to give a smaller reaction rate with respect to 
the fusion through the formation of the molecular complex: 
\begin{equation}
  t\mu + D_2 \rightarrow [(td\mu)dee]^* \rightarrow {\bf ^4He}^{++} + \mu + n + d+ 2e + Q.  \label{eq:dtmdee} \\
\end{equation}
(for review~\cite{bf,petitjean}).

Subsequently we determine the initial muon sticking probability, 
using the phase space distribution of the muon at the internal classical turning point.
The sticking probability is evaluated regardless of the difference of the two reactions 
eq.~(\ref{eq:dtmm0}) and (\ref{eq:dtmdee}) due to the fact that the Q-value is much larger 
than the binding energies of muonic atoms or molecules.
We suppose that $\omega_0$ is insensitive to the formation process of the muonic molecules. 
A distinctive feature of our study is 
that we do not assume the ground state of the muonic molecule
as the initial state of the muonic molecules, 
instead we use the initial state configuration by simulating 
the fusion process employing the imaginary time method~\cite{negele,bk1,bk2}.
As a consequence, in fact, the muon does not stick necessarily to the ground state of the alpha particle
and this fact plays an important role when we proceed to the stripping of the bound muon in the oscillating field.
The chaotic dynamics could prevent the muon from being lost in the $\mu$CF cycle due to the sticking.
It is achieved by utilizing the characteristic as a nonlinear oscillator of the trapped muon 
on the alpha particle.     
We draw an analogy between the muonic He ion in the present case and microwave-ionization of Rydberg 
atoms~\cite{mira,kv,krb}, where the driven electron in the highly excited hydrogen atom in a strong microwave 
electric field exhibits the chaotic dynamics and is ionized. 
Since highly excited states in the atom, with high quantum principle number $n$, are in the quasi-classical regime,
its stability can be explained in classical mechanics in terms of resonances. 
We carry out a numerical simulation by enforcing an oscillating field(linearly polarized, oscillatory electric field) 
on the system. 
This can be, likely, achieved by radiation of a coherent Synchrotron Orbital Radiation(SOR) X-ray experimentally. 
The oscillating 
force causes the resonance between the force itself and the oscillating motion of the muon around 
the alpha, especially when the driving frequency coincides with integer multiples
of the eigen frequency of the muonic helium. In other words the muon can be stripped by controlling the chaos of the system. 
In passing we mention that an attempt to apply magnetic fields for the purpose of stripping from excited
states of muonic helium has been proposed in~\cite{ms}.

This paper is organized as follows.
In Sec.~\ref{sec:form} we describe the theoretical framework 
of the CoMD for muonic molecule formation and following fusion process briefly.  
The relation between the enhancement factor and the chaotic motion of the muon is 
discussed in Sec.~\ref{sec:amsc}.
We develop in Sec.~\ref{sec:rad} a formula to estimate the initial $\alpha$-$\mu$ sticking 
probability($\omega_0$) and determine $\omega_0$.   
Sec.~\ref{sec:strip} is devoted to the discussion of a possibility of muon release.  
In Sec.~\ref{sec:sum} we summarize the paper and mention the future perspectives of this 
study. 

\section{Framework}
\label{sec:form}
The details of the framework of the CoMD is discussed 
in the references~\cite{pmb,kb-icfe,kb-ags}. 
In the following we sketch the framework briefly, by applying it to the case of the 
reaction~(\ref{eq:dtmm0}).
We assume the ground-state $t\mu$ as targets at the beginning of the collision. 
The ground-state 
muonic tritium configuration in the phase space is obtained using the CoMD 
approach~\cite{kb-ags}. Denoting the position of the particles($i=\mu, t$) in 
the phase space by (${\bf r}_i, {\bf p}_i$) and the relative distance and 
momentum between $\mu$ and $t$ 
by $r_{{\mu}t}=|{\bf r}_{\mu}-{\bf r}_t|$ and $p_{{\mu}t}=|{\bf p}_{\mu}-{\bf p}_t|$
respectively, 
the modified Hamilton equations for the muonic tritium with constraints are 
\begin{eqnarray}
  \label{eq:rt2}  
  \dot{{\bf r}}_{\mu} &=& \frac{{\bf p}_{\mu} c^2}{{\mathcal E}_{\mu}} 
  + \frac{1}{\hbar}
\frac{\lambda_{\mu}^H}{\xi_H}
  {r}_{{\mu}t}\frac{\partial {p}_{{\mu}t}}{\partial {\bf p}_{\mu} }, \\
  \label{eq:pt2} 
  \dot{{\bf p}}_{\mu} &=& -\nabla_{{\bf r}} U({\bf r}_{\mu})
  - \frac{1}{\hbar}
\frac{\lambda_{\mu}^H}{\xi_H}
  {p}_{{\mu}t}\frac{\partial {r}_{{\mu}t}}{\partial {\bf r}_{\mu} },
\end{eqnarray}
where we use relativistic kinematics;
${\mathcal E}_i= \sqrt{{\bf p}_i^2c^2+m_i^2c^4}$
and $U({\bf r}_{\mu})= -e^2/r_{\mu t}$   
is the potential of the muon.
The second term in the eqs.~\eqref{eq:rt2} and~\eqref{eq:pt2} represent 
the constraints:
the phase space density of two distinguishable particles should be always equal to 1 or less~\cite{bd},
i.e., the terms prevent the muonic tritium from collapsing. 
In terms of the phase space distance, this can be written as $r_{\mu t}p_{\mu t}\ge \xi_H \hbar$, 
where $\xi_H=1$ and the equal sign should be satisfied in the case of the ground state muonic atom. 
For this purpose
$\lambda^H_{\mu}$, the Lagrange multiplier for the Heisenberg principle,
is determined depending on the distance of the muon from the triton in the 
phase space ${r}_{\mu t}{p}_{\mu t}$. If ${r}_{\mu t}{p}_{\mu t}$
is (smaller)larger than $\xi_{H} \hbar$, 
$\lambda_i^H$ is a positive(negative) small finite number.
The approach gives the average binding energy of the ground state muonic tritium atom 
$BE_{t\mu}=-2.73$ keV. The value is in agreement with 
$BE_{H}(M_{\mu}/{M_e})=-2.71$ keV, where $BE_{H}$ is the binding energy of the hydrogen atom,
and $M_e$ and $M_{\mu}$ are the reduced masses of the the electronic and muonic atoms.

Using the obtained ground-state configuration as an initial state, we 
perform the numerical simulation of the fusion process~\eqref{eq:dtmm0} using  
\begin{eqnarray}
  \label{eq:rt3}  
  \dot{{\bf r}}_i &=& \frac{{\bf p}_i c^2}{{\mathcal E}_i}; \hspace*{0.5cm} 
  \dot{{\bf p}}_i = -\nabla_{{\bf r}} U({\bf r}_i)
\end{eqnarray}
for all the particles($i=t,d,\mu$). As the interaction we consider  
a modified Coulomb interaction $U({\bf r})=\sum_{j(\neq i)}q_i q_j/r_{ij}\times(1-e^{-br_{ij}})$, with 
$q_i$ and $q_j$ being the charges of the particles and $b=9500$\AA$^{-1}$. 

In order to treat the tunneling process in the framework of the molecular dynamics, we define 
the collective coordinates ${\bf R}^{coll}$ and the collective momentum ${\bf P}^{coll}$ as
\begin{equation}
  {\bf R}^{coll} \equiv {\bf r}_d-{\bf r}_t;   \hspace*{0.5cm}
  {\bf P}^{coll} \equiv {\bf p}_d-{\bf p}_t, 
\end{equation}
with ${\bf r}_t, {\bf r}_d$ ($ {\bf p}_t, {\bf p}_d$) being the coordinates(momenta) of 
the triton and the deuteron, respectively. 
To obtain the classical turning points, we first simulate the elastic collision. 
Subsequently we repeat the simulation in the tunneling region by
switching on 
the collective force, which is determined by ${\bf F}_d^{coll} \equiv \dot{\bf P}^{coll}$ 
and ${\bf F}_t^{coll} \equiv -\dot{\bf P}^{coll}$, to enter into imaginary time~\cite{bk1,bk2}.
We follow the time evolution in the tunneling region using the equations,
\begin{equation}
  \label{eq:rti}  
  \frac{d {\bf r}_{t(d)}^{\Im}}{d\tau}= \frac{{\bf p}_{t(d)}^{\Im}}{{\mathcal E}_{t(d)}}; \hspace*{0.5cm} 
  \frac{d {\bf p}_{t(d)}^{\Im}}{d\tau}= -\nabla_{{\bf r}} U({\bf r}_{t(d)}^{\Im}) -2{\bf F}^{coll}_{t(d)},
\end{equation}
where $\tau$ is used for imaginary time to be distinguished from real time.   
${\bf r}^{\Im}_{t(d)}$ and ${\bf p}^{\Im}_{t(d)}$ are position and momentum of the triton
(the deuteron) during the tunneling process respectively.   

We assume that the fusion process occurs at small impact parameters compared with 
the radius of the muonic atom, i.e., we carry out the simulation of the head on collisions.  
Under this assumption, 
the penetrability of the barrier is given by~\cite{bk1,bk2} 
\begin{equation}
  \label{eq:penet}
  \Pi(E)=\left(1+\exp\left(2{\mathcal A}(E)/\hbar\right)\right)^{-1},
\end{equation}
denoting the action integral ${\mathcal A}(E)$ as 
\begin{equation}
  {\mathcal A}(E)=\int_{r_b}^{r_a}{\bf P}^{coll}~d{\bf R}^{coll}
\end{equation}  
with $r_a$ and $r_b$ being the classical turning points.
The internal classical turning point 
$r_b$ is determined using the sum of the radii of the target and projectile nuclei.
Similarly from the simulation without muon, we obtain the penetrability of the bare 
Coulomb barrier $\Pi_0(E)$.
We choose the initial inter-nuclear separation 3 \AA.
This is much larger than the scale of the muonic tritium radius, which is of the order of 1.3 m\AA.  

\section{Enhancement of the cross section by the muonic Screening effect}
\label{sec:amsc}

We introduce the enhancement factor of the cross section by the bound muon $f_{\mu}$ 
\begin{equation}
  \label{eq:defenh0}
  f_{\mu}=\sigma(E)/\sigma_0(E),
\end{equation}
where $\sigma(E)$ and $\sigma_0(E)$ are the screened cross section and the bare cross section, 
respectively. We approximate Eq.~(\ref{eq:defenh0})  
by taking the ratio of the penetrabilities in the presence and in the absence of the muon:
\begin{equation}
  \label{eq:defenh}
  f_{\mu}=\Pi(E)/\Pi_0(E).
\end{equation}
In the following discussion, the enhancement factor is referred as an indicator of the regularity  
of the muonic motion~\cite{kb-cdf,kb-icfe}.  It plays a role of a sort of order parameter and it is 
determined through the values obtained in the numerical simulation. 
\begin{figure}[htbp]
  \centering
  \resizebox{65mm}{!}{\includegraphics{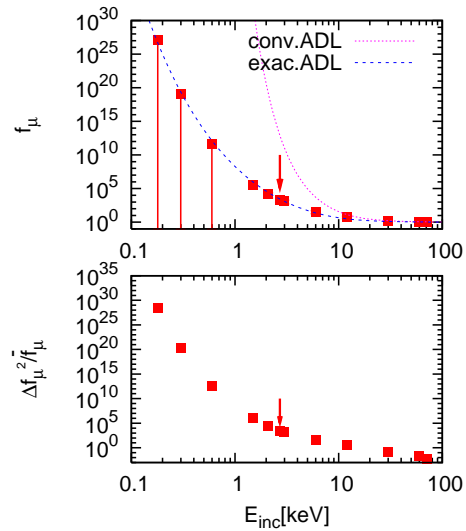}}
  \caption{Enhancement factor by the bound muon (top panel) and 
     $\Delta f_{\mu}^2/\bar{f}_{\mu}$ (bottom panel) as functions of the incident center-of-mass energy.
   The arrows in the figure indicate the point where total energy is zero.}
  \label{fig:EFm}
\end{figure}
In the top panel of Fig. \ref{fig:EFm} we plot $f_{\mu}$
as a function of the incident center-of-mass energy between the triton and the deuteron. 
From our simulation of the collisions using an ensemble of events,
we determine the average enhancement factor $\bar{f_{\mu}}$ and its variance: 
$\Delta f_{\mu}= [\bar{f_{\mu}^2}-\bar{f_{\mu}}^2 ]^{1/2}$.   
These are shown by squares and error-bars, respectively. 
Both the average $\bar{f_{\mu}}$ and its variance increase exponentially as the incident energy decreases. 
The dashed and dotted lines in the figure correspond to the enhancement factor $f_{\mu}$ in the 
conventional and exact adiabatic limit respectively~(See Appendix).  
The average $\bar{f_{\mu}}$ is in good agreement with the exact adiabatic limit with a screening 
potential $U_{\mu}=8.3$ keV.
This fact suggests that the enhancement mainly depends on the energies involved, i.e.,the energy 
difference of the initial and the final atoms. Thus the problem of fusion is somewhat independent 
on the detail of dynamics. 
Even if we had prepared the entrance channel through the formation of the molecular complex
eq.~(\ref{eq:dtmdee}), the actual fusion cross section would not differ from 
the one calculated here. In particular in the limit $E\ll U_{\mu}$ the fusion cross section can be 
written from eq.~(\ref{eq:defenh0}) 
and eq.(\ref{eq:cs}) as 
\begin{equation}
  \label{eq:cs2}
  \lim_{E\rightarrow 0}\sigma(E) = \lim_{E\rightarrow 0}\sigma_0(E+U_{\mu}) \rightarrow \frac{S(U_{\mu})}{U_{\mu}}e^{-2\pi \eta(U_{\mu})},
\end{equation}
where $\eta(U_{\mu})$ is the Sommerfeld parameter.

\begin{figure}[htbp]
  \centering
  \resizebox{95mm}{!}{\includegraphics{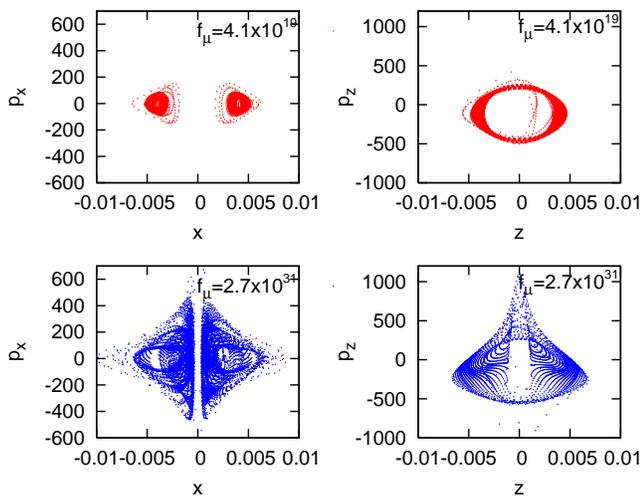}}
  \caption{Surface of section for 2 events, one has small $f_{\mu}$ (top panels) and the other has large $f_{\mu}$ (bottom panels),  
    on the $x$-$p_x$(left panels) and the $z$-$p_z$(right panels) planes at the incident c.o.m energy 0.18keV, in the atomic unit}
  \label{fig:POIx}
\end{figure}

In the bottom panel the ratio $\Delta f_{\mu}^2/\bar{f}_{\mu}$ versus incident energy is plotted. 
In the high energy limit the ratio approaches zero, i.e., the $f_{\mu}$ distribution 
becomes a $\delta$-function ($\Delta f_{\mu}=0$) and the average ${f_\mu}$ approaches 1:
there is no effective enhancement. 
In the low energy limit $\Delta f_{\mu}^2/\bar{f}_{\mu} \gg 1$, which implies that the system exhibits 
a sensitive dependence of the dynamics on initial conditions, i.e., 
occurrence of chaos. 
It is noteworthy 
that the slope of the ratio $\Delta f_{\mu}^2/\bar{f}_{\mu}$ changes 
at the ionization energy of the muonic tritium, which we indicated by the arrows in the figure.  
At this incident energy the total energy of the system is zero.
The total system might be unbound at the incident energies higher than this point, while 
the 3-body system is bound at lower energies.
We indeed verify the manifestation of chaos by plotting the Poincare surface of section
with respect to the enhancement factor for two events in Fig.~\ref{fig:POIx}. 
In the figure we show 
the surface of section for two selected events at the incident energy 0.18 keV 
on the $x$-$p_x$ plane~(FIG.~\ref{fig:POIx} left panels) and on the 
$z$-$p_z$ plane~(FIG.~\ref{fig:POIx} right panels), respectively. We choose the beam axis 
to coincide with the $z$-axis.  
At the incident energy 0.18 keV the average enhancement factor, $\bar{f_{\mu}}=$ 1.1 $\times 10^{27}$
as one can see in Fig.~\ref{fig:EFm}.  
In the top panels, with $f_{\mu}$= 4.1$\times 10^{19}$ ($\ll \bar{f_{\mu}}$) 
and the ratio of the external classical turning point in the presence of the muon($r_a^{\mu}$) to
the one in the absence of the muon($r_a^0$): $r_a^{\mu}/r_a^0=0.15$,   
the points show a map of a typical regular event. 
By contrast in the bottom panels, with $f_{\mu}$=
2.7$\times$10$^{31}$ ($>\bar{f_{\mu}}$) and the ratio of the classical turning points $r_a^{\mu}/r_a^0=0.06$, 
the points show the map of an irregular event: the points cover a large section of the map. 
The irregular muonic motion leads to smaller external classical turning point. As a consequence 
it gives larger enhancements factors opposed to the previous results in the case of the electron 
screening~\cite{kb-cdf,kb-icfe}, where the irregular(chaotic) events give smaller enhancement factors. 
This contradiction is accounted for the fact that the system is 
bound in the present case at low incident energies, while in the previous case even the lowest incident 
energy which was investigated is much higher than the binding energy of the electrons. 
Therefore the chaotic dynamics of the electrons causes to dissipate the kinetic energy between the target 
and the projectile and lowers the probability of fusion.    

We can easily deduce the reaction rate($\lambda^{dt}$) at the liquid 
hydrogen density($\rho_{LH}=4.25\times10^{22}cm^{-3}$) and at very low thermal energies $k_BT \ll U_{\mu}$, 
\begin{eqnarray}
  \lambda^{dt}&=&\lim_{T,E \rightarrow 0}\sigma(E)|\dot{\bf r}_t-\dot{\bf r}_d|\rho_{LH} \label{eq:rrate1} \\
  &=&\lim_{T,E \rightarrow 0}\sigma_0(E)f_{\mu}|\dot{\bf r}_t-\dot{\bf r}_d|\rho_{LH}. \label{eq:rrate2} 
\end{eqnarray}
Using the fact that the average enhancement by the muon is written as 
$\bar{f_{\mu}}=\sigma_0(E+U_{\mu})/\sigma_0(E)$ in terms of the screening potential $U_{\mu}$
and substituting the data in the NACRE compilation~\cite{nacre} for the bare reaction 
cross section $\sigma_0(U_{\mu})$ in eq.(\ref{eq:cs2}),
\begin{equation}
  \lambda^{dt}\sim\sigma_0(U_{\mu})\sqrt{\frac{2U_{\mu}}{M_{dt}}}\rho_{LH} \sim 2.8\times 10^4 [s^{-1}], \label{eq:rrate3} 
\end{equation}
where we denote the reduced mass of $d$ and $t$ as $M_{dt}$ and 
the charge symmetry in the t+d system has been taken into account as it is explained in the appendix. 
This value is in agreement with the result obtained with an independent approach~\cite{melezhik},
$\lambda^{dt}=0.5\times10^5 [s^{-1}]$. We, nevertheless, stress that this value is obtained using the average 
enhancement factor. As we have seen in Fig.~\ref{fig:POIx} fluctuations might be extremely large 
and there are events which have the $f_{\mu}$ more than 10$^4$ times larger than the average $\bar{f_{\mu}}$.     
Therefore the "in flight" $d-t$ fusion rate could fluctuate as well and become comparable to the inverse 
lifetime of the muon. 
We will discuss the role of fluctuations more in detail in a future work.

\section{Muon sticking probability}
\label{sec:rad}
We estimate the sticking probability of muons on the alpha particle in the exit channel: 
\begin{eqnarray}
  (td\mu)^+ &\rightarrow& {\bf ^4He}^{++} + \mu + n + Q \label{eq:anst} \\
            &\rightarrow& ({\bf ^4He}\mu)^{+} + n +Q. \label{eq:ast} 
\end{eqnarray}  
Recall that the Q-value is 17.59 MeV, while the energies involved both in the ``in-flight''
reaction (\ref{eq:dtmm0}) or with molecular complex formation eq.(\ref{eq:dtmdee}) is of the 
order of several eV, thus we expect that the actual entrance channel for fusion is irrelevant
to the following dynamics of tunneling, fusion, decay and sticking of the muon to the alpha particle.

The muon remains bound (eq.~\eqref{eq:ast}), if   
the binding energy of the muon on an alpha particle:
\begin{equation}
  BE_{\alpha\mu}=\frac{M_{\alpha\mu}}{2}|\dot{{\bf r}}_{\mu}-\dot{{\bf r}}_{\alpha}|^2-\frac{2e^2}{|{\bf r}_{\mu}-{\bf r}_{\alpha}|} \label{eq:beam}
\end{equation}
is negative, in the center-of-mass system of the muon and the alpha particle. 
We denote the reduced mass of $\mu$ and $\alpha$ as $M_{\alpha\mu}$.
The effect of the finite nuclear mass must be taken into account, because a muon is 206.8 times heavier 
than an electron.   
From this condition, $BE_{\alpha\mu}\le 0$, we deduce the following equation for the 
angle $\theta$ between $\dot{{\bf r}}_{\mu}$ and $\dot{{\bf r}}_{\alpha}$.  
\begin{eqnarray}
  \cos{\theta}
  \ge\frac{\frac{M_{\alpha\mu}}{2}(|\dot{{\bf r}}_{\mu}|^2+|\dot{{\bf r}}_{\alpha}|^2)-\frac{2e^2}{|{\bf r}_{\mu}-{\bf r}_{\alpha}|}}
  {M_{\alpha\mu} |\dot{{\bf r}}_{\mu}| |\dot{{\bf r}}_{\alpha}|} \equiv g  \label{eq:rhs}
\end{eqnarray}
The condition Eq.~\eqref{eq:rhs} is fulfilled when $g$, the r.h.s of the equation, is 
equal to 1 or less and for 
the solid angle $\Omega=2\pi (1-g)$ [steradian] in the 3-dimensional space. 
We can therefore estimate the sticking probability by $\Omega/4\pi$, if $g \le 1 $.  
We point out that $|\dot{{\bf r}}_{\alpha}|$ in the equation is written as a function of the decay Q-value:
\begin{equation}
  \label{eq:ral}
  |\dot{{\bf r}}_{\alpha}|= \frac{m_n}{m_{\alpha}+m_n}|\dot{{\bf r}}_{\alpha n}|= \frac{m_n}{m_{\alpha}+m_n}\sqrt{2Q/M_{\alpha n}},
\end{equation}
where $\dot{{\bf r}}_{\alpha n}=\dot{{\bf r}}_{n}-\dot{{\bf r}}_{\alpha}$ is the relative velocity between the $\alpha$ and the neutron
and $M_{\alpha n}$ is their reduced mass.
In particular, the sticking probability can be estimated easily in some limiting cases:\\
CASE 1. if $|\dot{{\bf r}}_{\mu}|$ is about $|\dot{{\bf r}}_{\alpha}|$, 
\begin{equation}
  \label{eq:g1}
  g\sim 1-\frac{2e^2}{|{\bf r}_{\mu}-{\bf r}_{\alpha}|}\times|\dot{{\bf r}}_{\mu}|/M_{\alpha\mu} \le 1
\end{equation}
Furthermore, in addition, we assume that the muon is bound in 
the ground state of the {\bf $^5$He} at the moment of the fusion(adiabatic limit), i.e., 
$-\frac{2e^2}{|{\bf r}_{\mu}-{\bf r}_{\alpha}|}=-10.942\times2.0$ keV, 
$g$ is estimated to be 0.89 and thus the sticking probability is 5.6 \%. 
In passing we mention that $g$ is  
0.352 for the reaction $d+d \rightarrow ^3${\bf He} $+n+3.268$ MeV under the same assumptions with CASE 1. 
We deduce 32.4 \% of the muon sticking probability in this case.  
In the case of the reaction $t+t \rightarrow \alpha +n+n+11.33$ MeV~\cite{matsuzaki}, if we assume 
that two neutrons bring away the maximum energy 9.44 MeV, we can estimate $g=$ 0.84 and 
the sticking probability= 7.84 \% for this reaction. \\
CASE 2. if $|\dot{{\bf r}}_{\mu}|$ is much smaller than $|\dot{{\bf r}}_{\alpha}|$,
\begin{equation}
  \label{eq:g2}
  g\sim \frac{1}{2}
  \frac{|\dot{{\bf r}}_{\alpha}|}{|\dot{{\bf r}}_{\mu}|} \ge 1,
\end{equation}
where we assume $-\frac{2e^2}{|{\bf r}_{\mu}-{\bf r}_{\alpha}|}\sim 0.0$. This means 
there is no sticking probability in this case. \\

\begin{figure}[htbp]
  \centering 
  \resizebox{75mm}{!}{\includegraphics{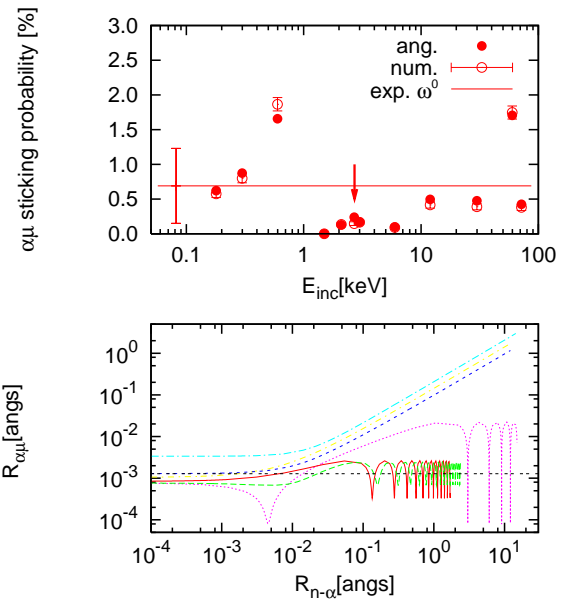}}
  \caption{Incident energy dependence of the sticking probability of the muon on the $\alpha$ particle.
  The statistical error is shown by error bars, otherwise it is within the size of the points in the 
  figure. (top panel)
  Distance between the muon and the alpha particle as a function of the inter-nuclear separation (bottom panel)}
  \label{fig:STP}
\end{figure}

We remind that muons can have higher velocity components in the quantum mechanical system. 
Turning to the case of our simulations, the above velocity for the bound muon is obtained as the average velocity over 
the ensemble of events. 
By inspecting each event, one could find some events which satisfy the condition $g < 1$.
Indeed one of the two events shown in the top panels in Fig.~\ref{fig:POIx}, which is regular, has  
$g= 0.93 (< 1.0)$ therefore the sticking probability of this event itself is not zero (3.8 \%). 
While the other in the bottom panels has $g= 1.9 (> 1.0)$, the sticking probability is zero. 
In the same way we calculated $g$ for all the events which are created in our simulation.
The resulting sticking probability of the muon on the $\alpha$ particle is shown with filled circles 
in the top panel of Fig.~\ref{fig:STP} as a function of the incident energy of the collision. 
At the same time, we carry out the simulation of the exit channel 
by creating 20000 events(randomly chosen directions of outgoing particles).
The time integration of the equation of motion, Eq.~\eqref{eq:rt3} 
for all the particles($i=\alpha,n,\mu$) is performed using Predictor-corrector 
integration scheme.   
We distinguish the muon sticking event~\eqref{eq:ast} from 
the release event~\eqref{eq:anst} by monitoring 
the binding energy of the muon on alpha particles Eq.~\eqref{eq:beam} 
and the radius of the muonic ion, 
${\mathcal R}_{\alpha\mu}=|{\bf r}_{\mu}-{\bf r}_{\alpha}|$. 
We count the events where 
$BE_{\alpha\mu}$ and ${\mathcal R}_{\alpha\mu}$ maintain to be negative and small
respectively, 
up to the point where the alpha is distant enough from the neutron.

In the bottom panel in Fig.~\ref{fig:STP} typical trajectories of the distance between the muon and the alpha 
particle are shown as a function of the inter-nuclear separation. 
Among the 6 curves shown in the figure, 
3 curves, which show oscillational behaviors, correspond to the sticking events.
While 3 other curves increase monotonically after $R_{n \alpha}$ exceeds 0.01 \AA.  
The horizontal straight line in the figure indicates the size of the radius of the ground 
state muonic {\bf He} atom.  
The obtained sticking probability is shown with open circles with error-bars in the top panel 
in Fig.~\ref{fig:STP}. 
First, as one can see clearly, the result of numerical simulation agrees with the sticking probability 
which is calculated considering the solid angle $\Omega$. 
In the figure we plot the sticking coefficient obtained from the direct measurement of 
$\omega_0$~\cite{davies}  
with a solid line with an error bar.
The resulting sticking probability range nearby the experimental value as a function of the incident energy
except for several points, which have zero and relatively large sticking probability.
This fact supports our assumption that $\omega_0$ is insensitive to the formation process of the muonic molecules.
We mention especially that the incident energies at which the sticking probability becomes zero is slightly below 
than the ionization energy of the muonic tritium.  
 
\section{Muon Stripping} 
\label{sec:strip}
Before concluding the paper, we would like to suggest that the stuck muon is possibly stripped
from the alpha particle, by enforcing a linearly polarized oscillatory electric wave on the system.~\cite{ms}   
The periodic motion of the stuck muon can be expressed in terms of nonlinear oscillations.  
For a nonlinear oscillator the oscillating driving force, i.e., linearly polarized, in the direction $z$,  
oscillating field of frequency $\gamma$ and a peak amplitude $F$ of the field,  
\begin{equation}
  \label{eq:efosc}
  z F \sin(\gamma t),
\end{equation}
causes the resonance between the force itself and the oscillating motion of the muon at 
driving frequencies which are integer multiples of the fundamental frequencies of the muon.
\begin{figure}[h]
  \centering
  \resizebox{85mm}{!}{\includegraphics{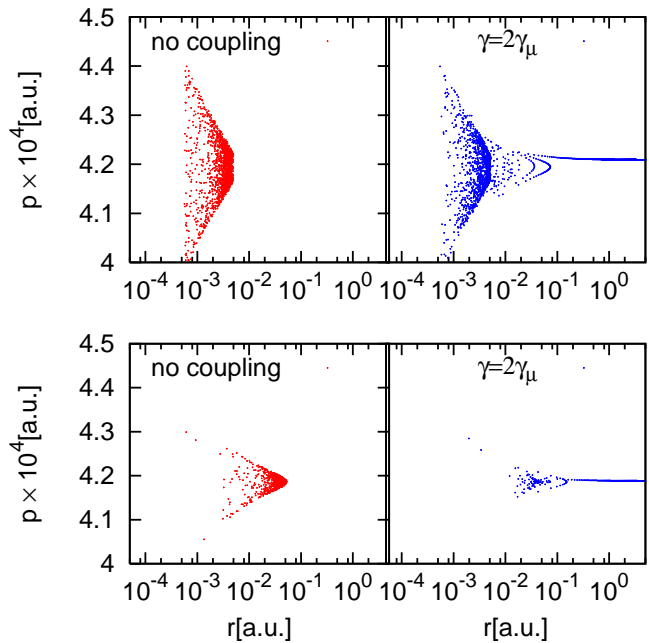}}
  \caption{Time $T$ map of a stuck muon on the $r$-$p$ plane(left panels). 
           One with the external oscillating force (right panels) with frequency $\gamma$, 
           both in the atomic unit}
  \label{fig:POIr}
\end{figure} 
The same concept has been applied in the ionization of the Rydberg atoms in a microwave field. 
There the highly excited atom, Rydberg atom, is prepared by laser excitation.    
In the present case of muons, the muonic {\bf He} is not stuck necessarily in its ground state. So that 
the muon can be ionized directly from one of such an excited state by the radiation of an electrostatic wave, 
otherwise the muon is, at first, prompted to an excited state and then ionized. 
In either cases it will be achieved using X-rays, since   
the fundamental frequencies for the ground and the first excited state of the muonic {\bf He} ion
corresponds to 0.11nm and 0.44nm, respectively, in terms of the wave length. 
Such a X-ray is available from Synchrotron Orbital Radiation(SOR) facilities.
In our numerical simulation, instead of the discontinuous frequencies, we get 
the proper frequency of the stuck muon for each event. The frequency is obtained by 
Fourier transform of the oscillation and with thus obtained frequency $\gamma_{\mu}$ we perform 
the simulation of stripping with above external force.  
In Fig.~\ref{fig:POIr} 
we show the time $T$ map of the oscillational motion of the muon on the $r-p$ plane, 
where $r$ is relative distance between the muon and the alpha and $p$ is its conjugate momentum. 
The time $T$ map is obtained by sampling the points in the phase space at discrete times~\cite{strogatz} 
\begin{equation}
  \label{eq:t}
  t=nT \  (T= 2\pi/\gamma, n=1,2,\cdot\cdot\cdot).
\end{equation}
We follow 1000 cycles of the driven oscillation in our simulation.
We choose two sticking event which are shown in the bottom panel in the figure~\ref{fig:STP};
one has a smaller amplitude, a tightly bound state(top panels), and the other has a 
larger amplitude, a loosely bound state(bottom panels).    
The left panels show the map of the stuck muon without external force.
The map remains in a limited manifold around $r=0$. 
The right panels show the case with an external force with driven frequency $\gamma=2\times\gamma_\mu$, 
where $\gamma_\mu=BE_{\alpha\mu}/\hbar$ is the angular frequency of the muon in the muonic helium
ion, i.e., we are investigating the 2:1 resonance. 
In the case of the tightly bound muon, the muonic atom is excited in a loosely bound state and then ionized.  
While the tightly bound muon is ionized in one step.  
The muonic atom in the external oscillating field 
is captured into the resonance and ionized due to its stochastic instability~\cite{PhysRevE.65.046230}.     
One can see clearly that the muon is expelled from the Helium with the external oscillational force
with the corresponding frequency of the unperturbed system.   
We point out that the $\mu dt$ molecule is {\it not} destroyed by the external force with same frequency which we 
used in the above discussion.

\vspace*{1cm}

\section{Conclusions and Future Perspectives} 
\label{sec:sum}
In this paper we discussed the alpha-muon sticking coefficient in muon-catalysed "in flight" d-t fusion. 
We performed numerical simulation by the Constrained Molecular Dynamics model. 
Especially the influence of muonic chaotic dynamics 
on the sticking coefficient is brought into focus.
The chaotic motion of the muon affects not only the fusion cross section but also 
the $\mu-\alpha$ sticking coefficient. The irregular(chaotic) dynamics of the bound muon leads to larger enhancements 
with respect to regular systems because of the reduction of the tunneling region. Moreover they give smaller sticking 
probabilities than those of regular events. 

We proposed a method to strip the stuck muon from the alpha particle by exposing the system 
in the X-ray radiation field. 
Its numerical experiments have been performed under an oscillating external force with the driving 
frequency twice as high as the angular frequency of the stuck muon and the muon has been released 
successfully with the selected frequency. 
By utilizing the chaotic dynamics one can prevent the muon from being lost in the $\mu$CF cycle 
due to sticking.

Based on these results, in our future study, we will develop a theory to investigate 
the temperature dependent phenomena, including the sticking probability and the muon cycling rate, 
which are reported by experimentalists~\cite{bom,kawamura1,kawamura2}.
Further quantitative analysis of the muon stripping with the oscillating force should be undertaken.

\vspace*{1cm}

\begin{center}
{\bf Acknowledgments}  
\end{center}
The authors acknowledge Prof. N. Takigawa for useful comments. 
This work was partly carried out during a short term stay of one of us(S.K) at University of Ferrara in Italy.  
She is grateful to Prof. G. Fiorentini and Prof. B. Ricci for suggestive discussions and their hospitality.     

\appendix*

\section{Enhancement factor in the adiabatic limit}
\label{sec:efad}
In the section ~\ref{sec:amsc} we introduced the enhancement factor of the cross section 
by bound muon in terms of the barrier penetrability. 
In the case of the electron screening, one often assumes that the effect of the screening 
can be represented by a constant shift, $U_{\mu}$, of the potential barrier 
and replaces eq.(\ref{eq:defenh0}) by   
\begin{equation}
  \label{eq:fmu2}
  f_{\mu}=\frac{\sigma_0(E+U_{\mu})}{\sigma_0(E)}.
\end{equation}
$U_{\mu}$ is called screening potential. 
$\sigma_0(E)$ can be rewritten in terms of the S-factor $S(E)$ and Sommerfeld parameter
$\eta(E)$,
by writing down the incident energy dependence of the barrier crossing rates explicitly~\cite{clayton},
\begin{equation}
  \sigma_0(E)=\frac{S(E)}{E}e^{-2\pi \eta(E)}. \label{eq:cs}
\end{equation}
In the limit of $U_{\mu} \ll E$, $f_{\mu}$ is approximated by 
\begin{equation}
  \label{eq:fconv}
  f_{\mu}=\exp{\left[\pi \eta(E) \frac{U_{\mu}}{E}\right]}. 
\end{equation}
The merit of this conventional formula is that 
one can easily estimate the upper limit 
of the enhancement by using the adiabatic approximation   
in the framework of the Born-Oppenheimer approximation. 
In the present case of the muonic tritium target,
reflecting the charge symmetry in the t+d, $f_{\mu}$ is obtained by assuming equally 
weighted linear combination of the lowest-energy ``gerade'' and ``ungerade'' wave function for the muon.~\cite{kb-cdf}
\begin{equation}
  \label{eq:fconv2}
  f_{\mu}=\frac{1}{2}(e^{\pi \eta(E) \frac{U_{\mu}}{E}}+e^{\pi \eta(E) \frac{U_{\mu}^{(u)}}{E}}). 
\end{equation}
it is given by substituting $U_{\mu}=BE_t-BE_{^5{\bf He}} \sim 8.3$ keV with 
$BE_t$ and $BE_{^5{\bf He}}$ being the binding energy of the ground state muonic tritium and the ground state 
muonic {\bf $^5$He} respectively and $U_{\mu}^{(u)}=39.6$ eV. 
The point is, however, that the procedure is justified only in the limit where $U_{\mu}$ is much smaller than $E$. 
$U_{\mu}$ is much larger than the low incident energies of our interest.
The dotted curve in Fig.~\ref{fig:EFm}, which corresponds to Eq.~(\ref{eq:fconv2}), indeed overestimates 
the enhancement with respect to the exact formula:     
\begin{equation}
  \label{eq:ff}
  f_{\mu}=\left(1+\frac{U_{\mu}}{E}\right)^{-1}
  e^{-2 \pi \eta(E)\left(\frac{1}{\sqrt{1+U_{\mu}/E}}-1\right)},
\end{equation}
where we approximated nothing but $S(E+U_e)\sim S(E)$.
To apply this formula in the present case again one needs to take a linear combination of the muonic molecular states.
This is taken into account and gives 
the dashed curve in Fig.~\ref{fig:EFm}.
The curve is in accord with the average enhancement factor obtained from our simulation. 


\bibliography{lamst.bib}
\end{document}